\documentclass[10pt,conference]{IEEEtran}
\IEEEoverridecommandlockouts

\newcommand{\qq}[1]{\textit{``#1''}} 

\usepackage{cite}
\usepackage{amsmath,amssymb,amsfonts}
\usepackage{algorithmic}
\usepackage{graphicx}
\usepackage{textcomp}
\usepackage{xcolor}
\usepackage{tabularx}
\usepackage{booktabs}
\def\BibTeX{{\rm B\kern-.05em{\sc i\kern-.025em b}\kern-.08em
    T\kern-.1667em\lower.7ex\hbox{E}\kern-.125emX}}

\begin{document}

\title{Hybrid Work in Agile Software Development: Recurring Meetings\\
\thanks{This research has been supported by the Finnish Work Environment Fund, under grant \#240172.}
}

\author{\IEEEauthorblockN{Emily Laue Christensen}
\IEEEauthorblockA{\textit{Software Engineering} \\
\textit{LUT University}\\
Lahti, Finland \\
emily.christensen@lut.fi}
\and
\IEEEauthorblockN{Maria Paasivaara}
\IEEEauthorblockA{\textit{Software Engineering} \\
\textit{LUT University}\\
Lahti, Finland \\
maria.paasivaara@lut.fi}
\and
\IEEEauthorblockN{Iflaah Salman}
\IEEEauthorblockA{\textit{Software Engineering} \\
\textit{LUT University}\\
Lahti, Finland \\
iflaah.salman@lut.fi}
}

\maketitle

\begin{abstract}
The Covid-19 pandemic established hybrid work as the new norm in software development companies. In large-scale agile, meetings of different types are pivotal for collaboration, and decisions need to be taken on how they are organized and carried out in hybrid work. This study investigates how recurring meetings are organized and carried out in hybrid work in a large-scale agile environment. We performed a single case study by conducting 27 semi-structured interviews with members of 15 agile teams, product owners, managers, and specialists from two units of Ericsson, a multinational telecommunications company with a ``2 days per week at the office'' policy. A key insight from this study is that different types of meetings in agile software development should be primarily organized onsite or remotely based on the meeting intent, i.e., meetings requiring active discussion or brainstorming, such as retrospectives or technical discussions, benefit from onsite attendance, whereas large information sharing meetings work well remotely. In hybrid work, community meetings can contribute to knowledge sharing within organizations, help strengthen social ties, and prevent siloed collaboration. Additionally, the use of cameras is recommended for small discussion-oriented remote and hybrid meetings.

\end{abstract}

\begin{IEEEkeywords}
hybrid work, large-scale agile software development, software team, meeting, community of practice, workspace.
\end{IEEEkeywords}

\maketitle

\section{Introduction}
The Covid-19 pandemic moved software professionals to their home offices, and their employers have found it difficult to get them back to the office due to the benefits of remote work, such as better work-life balance \cite{smite2022half}. After the pandemic, hybrid software development, defined by Conboy et al. \cite{conboy2023future} as a setting where: \qq{some team members work mostly or completely from home, others mostly or completely from the traditional office, and others in some combination of the two—not quite distributed and not quite co-located but, rather, individuals working from anywhere and touching base with the office intermittently}, soon became the norm.

As the topic of hybrid work in software engineering is new, there is not yet much systematically collected research data and results available. Existing studies focus mainly on remote work during the pandemic (e.g., \cite{yang2022effects, miller2021your, muller2023challenges}), but less so on post-pandemic hybrid work (e.g., \cite{jaspan2023developer, tkalich2023pair, de2023post}). A recent systematic mapping study on hybrid work in agile software development \cite{khanna2024hybrid} found only twelve primary studies on the topic, which shows a severe lack of research. Moreover, Conboy et al. \cite{conboy2023future} urge the need for future research on hybrid work in software engineering. With this study we aim to start filling that gap.

The most commonly used software engineering methodology, agile software development \cite{agilereport}, is based on face-to-face communication inside an agile team \cite{beck2001agile}. Thus, hybrid work in agile development necessitates rethinking how to organize communication and collaboration. For instance, Šmite et al. \cite{smite2022future} argue that when software teams are fully aligned, i.e., they \qq{move between the office and the remote modes in an aligned fashion}, this can help to avoid challenges associated with hybrid work and level the playing field experience. 

Large-scale agile requires even more consideration because collaboration and communication between several agile teams, other stakeholders, and the rest of the organization needs to be planned. Typically, communication in large-scale agile is based both on ad hoc communication and on regular meetings, for example, daily stand-up meetings \cite{stray2016daily, stray2018daily}. In this study, we concentrate on one of the main forums for communication and collaboration in hybrid work in a large-scale agile environment, i.e., recurring meetings. These are scheduled meetings that usually occur at regular intervals, in contrast with ad hoc meetings, which are carried out sporadically. In the context of our case organization, a global telecommunications company --- Ericsson, we studied our research question through 27 semi-structured interviews: \textit{``How are recurring meetings organized and carried out in hybrid work in agile software development?''} 

\section{Related Work}
Challenges in communication and collaboration among software professionals were common themes identified in several studies carried out during the Covid-19 pandemic (e.g., \cite{rodeghero2021please, yang2022effects, cao2021large}), due to the sudden shift to remote work. For instance, Yang et al. \cite{yang2022effects} found that remote work caused more static and siloed collaboration among a network of Microsoft employees and suggested that hybrid work policies may have similar effects. Whereas, Rodeghero et al. \cite{rodeghero2021please} noticed that developers often kept their cameras off during remote meetings, which made collaboration even more difficult. Cao et al. \cite{cao2021large} found that multitasking was a more prevalent behavior in recurring remote meetings, than in ad hoc meetings, with often negative effects, such as distraction. 

However, only a few studies focus on meetings in hybrid work, in agile software development. The systematic mapping study by Khanna et al. \cite{khanna2024hybrid} found only three studies that research the topic, and two additional studies that discuss meetings in connection with hybrid work. Next, we report the main findings of these studies.

Hybrid scrum meetings were investigated by both Buyukguzel and Mitchell \cite{buyukguzel2023progressivity}, and Buyukguzel and Balaman \cite{buyukguzel2023spatial}. The findings of these studies provide insights into how meeting participants establish and maintain progressivity, i.e., the continuity of work, by moving activities forward when disruptions occurred \cite{buyukguzel2023progressivity}, and how they co-construct space, for example, by ensuring that participants at the office are within the screen radius, so they are visible and audible to remote participants \cite{buyukguzel2023spatial}. Unscheduled meetings in hybrid work were explored by Sporsem et al. \cite{sporsem2022unscheduled}, to determine how developers maintain ad hoc communication. The findings showed that virtual rooms (e.g., on Discord or Zoom) help determine the need for unscheduled meetings, while Slack channels with fewer members offer a safe and free discussion environment, which also led to unscheduled meetings \cite{sporsem2022unscheduled}.

Although meetings are not the focus of \cite{sporsem2022coordination} and \cite{wang2022co}, both of these studies of agile software teams revealed interesting insights on the topic which we briefly report here, and discuss further in connection with our findings in Section \ref{sec:discussion}. Firstly, in their study of workplace design, Wang et al. \cite{wang2022co} found that teams need to agree on predefined onsite days for highly collaborative work, like product backlog refinement meetings. The authors also recommend coordinating and leveraging onsite time deliberately, while limiting the number and frequency of hybrid meetings, as these meetings can defeat the purpose of onsite work hours, which should be utilized for in-person collaboration and fostering social interaction.

Similarly, in their study of ``work-from-anywhere'' coordination strategies, Sporsem and Moe \cite{sporsem2022coordination} found that teams avoid filling up their calendars with meetings during onsite office days, to enable collaborative work, so large meetings and retrospectives were exclusively held virtually on days where participants work from home. 

In contrast to the aforementioned studies which focus primarily on remote meetings and on details within hybrid meetings specifically, we explore recurring meetings in hybrid work in a large-scale agile environment, and gather data on numerous types of meetings at both unit and team levels, as well as in management. Thus, our study aims to provide insights into how a whole set of meetings in hybrid work in a large-scale agile environment can be organized and carried out, thereby contributing to addressing the current scarcity of literature in this domain.

\section{Research Design}

To investigate recurring meetings in hybrid work in agile software development, we conducted an exploratory single case study following the guidelines of Yin \cite{yin2018case} and Runeson and Höst \cite{runeson2009guidelines}. A protocol for the case study was created and continuously updated throughout the research process \cite{brereton2008using}. 

\subsection{The Case Company and Context}

The case company is Ericsson, a multinational telecommunications company that employs about 99,000 people worldwide. On the company level, Ericsson transitioned to agile in late 2012. The site investigated in this study is the Finnish R\&D site that concentrates on product and network security, cloud, 5G, and 6G research \cite{ericsson}. This site started its lean and agile transformation in 2009 \cite{paasivaara2018large}. The site employs around 700 professionals and applies a hybrid work policy where most employees are expected to work at the office two days per week. We investigated two units of the same site, which each implement the hybrid work policy in different models: Unit 1, where the 65 employees are expected to work at the office on Tuesdays and Thursdays, and Unit 2 where the two office days are flexible for the 70 employees. Unit 2 also closely collaborates with other sites in Europe and Asia, but our study only focuses on the two units in the Finnish site.

The hybrid work models were in an experimental phase at the beginning of our research. The R\&D managers were interested in learning how the different hybrid work models and work arrangements of the teams impact collaboration and agile practices. Teams have the autonomy to decide which agile practices and events to apply in their work. In Unit 1, the six teams follow the Scrum framework quite closely. Whereas, the agile development process models followed by the nine teams in Unit 2 are more varied and combine aspects of Scrum and Kanban. None of the teams from the two units have a specific member acting as a Scrum master. However, most of the teams in Unit 2 implement a two-week rotation for their members to serve as team coaches. The agile teams varied in size: eight teams have between 4--6 members, six teams have between 7--9 members, and one team has 13 members. 

The software project in Unit 1 has been in development for a couple of years. Whereas, in Unit 2 product development started over 20 years ago with the traditional waterfall way of working having system, design, and verification phases separately. Since early 2010 these were combined, and large-scale agile development was started \cite{hallikainen2011experiences}. The product itself has evolved throughout these years from native to virtual and cloudified versions. The interviews took place just before the launch of fixed team areas and workplaces for the first time after almost full remote work due to the pandemic. The two units were in the process of redesigning their office workspace to better accommodate these changes and the hybrid work models. The management was therefore interested in assessing the layout of the office premises to determine the value of fixed seating and other areas of the workspace, such as the meeting rooms. These interests aligned well with our research objectives, providing a compelling case for our study. In this paper, we narrow the scope to investigate recurring meetings.

To plan the research with the case company, a meeting was arranged between the researchers and several site managers who presented the organization of the two units chosen for this study, the policy and models for hybrid work, the product domains, and the tentative future layout of the office workspace. The researchers also visited the office premises to observe the current work environment of the two units. Additionally, the meeting contributed to formulating our interview guide, which can be viewed in the supplementary materials\footnote{Supplementary materials: https://doi.org/10.6084/m9.figshare.25761945.v1\label{fn}}. 

Finally, the managers aided in sampling participants for interviews per the requirements of the researchers. To ensure that interviewees represented varying modes of work, roles, and teams in the company, we used purposive sampling, by asking the managers to suggest volunteers based on the following criteria: 1) one member from each agile team, in both units, 2) people outside of the teams who worked in varying roles, and 3) people with varying office presence, e.g., some who often worked from the office, others who rarely worked at the office, fully remote people, and those working according to the ``2 days per week at the office'' policy.

\subsection{Data Collection}
\label{sec:dataCollection}
Qualitative data was collected via semi-structured interviews with 27 employees located in Finland. The employee's interview participation was voluntary and consent-based. Ten participants were from Unit 1 while the remaining 17 worked in Unit 2. Both male and female interviewees participated and their roles included members from all 15 agile teams, product owners (PO), managers, and specialists. An overview of the interviewee roles, years of experience in the company, and the participant ID's (P1--P27) can be seen in Table \ref{tab:demographics}.

The interviews were carried out via Microsoft (MS) Teams between November 2023 and February 2024 by one to three interviewers in English. Most interviews were conducted by two interviewers. At the start of the interviews, the participants were informed about the objectives of the study, how the results would be used, and the precautionary steps to maintain confidentiality. We also included this information in the consent form which all the participants signed. We then obtained their explicit verbal permission to record the interviews. Each interview lasted 60 minutes on average. All the interviews were video recorded via MS Teams with an automatic transcription-generating feature. 

The first agile team member we interviewed was P3 [Team B]. After this the interview guide was refined by the research team to ensure more details were collected about how the meetings were carried out, such as the exact location when onsite, who facilitated, and camera usage. Given that the interviews were semi-structured, some of the meeting descriptions provided by the following interviewees were still slightly more detailed than others. Follow-up emails were therefore sent to 13 interviewees in April 2024 to inquire about a few particulars that were missing, for example, the duration of certain meetings. Ten interviewees provided responses to these emails which were added to their respective transcriptions. We did unfortunately not obtain a response from P3 [Team B], so the meeting descriptions provided by that interviewee lack some details, as can be seen in Figure \ref{fig:teammeetings}.

\subsection{Data Analysis}
\label{sec:dataAnalysis}

We adopted a codebook thematic analysis approach to analyze and code our data \cite{braun2021one}. The process began with manually editing the MS Teams generated interview transcripts to improve the accuracy and familiarize with the data, and ideas for coding were noted down. The first author generated an initial codebook containing 16 primary codes with their descriptions based on the notes taken, the study objectives, and the interview guide. These primary codes were limited to different \textit{types of meetings}, the form of the meetings (\textit{hybrid, remote, and onsite}), and \textit{tools} used during meetings. A deductive approach was then used to identify data items that matched those 16 codes in three interview transcripts using NVivo R1. The codebook was then discussed in the research team and 10 additional codes were added based on the data from the first three interviews.

The first author then began a recursive process where the codes were identified in the data of all the interviews. As each code was added, the interviews which had been previously analyzed were reanalyzed to include each new code until the process was complete. An example of coding from one of the interviews is provided in the supplementary materials\footref{fn}. The codes were then sorted into themes which were further reviewed and refined, then defined and named jointly by the research team, resulting in the final codebook which contained 43 codes with distinctive descriptions\footref{fn}. Finally, the quotations from the interviews reported in this study were cleaned to remove filler words and repetitions, and anonymized to prevent the identification of participants. Every quotation is followed by the participant ID and either the agile team ID (A--O), for example, P9 [Team F], or the interviewee role, for example, P1 [Manager].

 \begin{table}[t]
\caption{Interviewee Demographics}
\label{tab:demographics}
\centering
\footnotesize
\begin{tabular}{ p{2.8cm} p{2.2cm} p{2.6cm} } 
 \toprule
\multicolumn{1}{l}{\textbf{Role}} &
\multicolumn{1}{l}{\textbf{Experience}} &
\multicolumn{1}{l}{\textbf{Participant ID}} \\ [0.3ex] 
 \midrule
Agile team member (16) & 4 months--34 years & P2--P4, P7--P9, P14, P15, P18, P20--P26 \\ 
Product owner (3) & 2--23 years & P6, P12, P13 \\ 
Manager (5) & 10--38 years & P1, P11, P16, P17, P27 \\ 
Specialist (3) & 4--26 years & P5, P10, P19\\  
 \bottomrule
\end{tabular}
\end{table}

\subsection{Feedback Sessions}
\label{sec:feedback}
Providing feedback and presenting analyses to participants is a critical step for maintaining trust and ensuring that the interpretation of the findings is reliable \cite{runeson2009guidelines}. So, after the interviews, feedback sessions were carried out with each unit where the research team summarized the main findings from the interviews in a 25 minute presentation, followed by a 10 minute Q\&A and discussion with the attendees. The feedback sessions were arranged at the earliest possibility after the data collection, as the management wished to use the findings and relevant discussions for informed decision-making in the near future. 

Both sessions were open to all the employees of the units, so all interviewees were invited, and the sessions were arranged virtually via MS Teams. Unit 1 had 73 people attending the session virtually, as it was held on a remote office day. Unit 2 had 42 people attending virtually, whereas about 10 people participated in-person from the unit's common sofa area. During the sessions, we recognized most of the interviewees, but unfortunately could not verify if all were present. The attendees found the results to be insightful and felt that they provided a good overview of the ways of working in the units, within the established hybrid work models. During the Q\&A and discussion the attendees inquired how the findings from the units compare in general with the industry, and some specific work practices were discussed in more detail, for example, how to avoid interrupting other employees, and the use of cameras during meetings. The outcome of the sessions was only used to validate our findings, and no corrections to our results were suggested.

\section{Results}

In this section, we present the results. We describe first the office presence of the interviewees and teams, and the office workspace and tools used during and in connection with meetings, and follow with descriptions of how the various recurring meetings are organized and carried out.

\subsection{Office Presence}

\begin{figure}[t]
    \centering
    \graphicspath{ {./images/} } \includegraphics[width=0.40\textwidth]{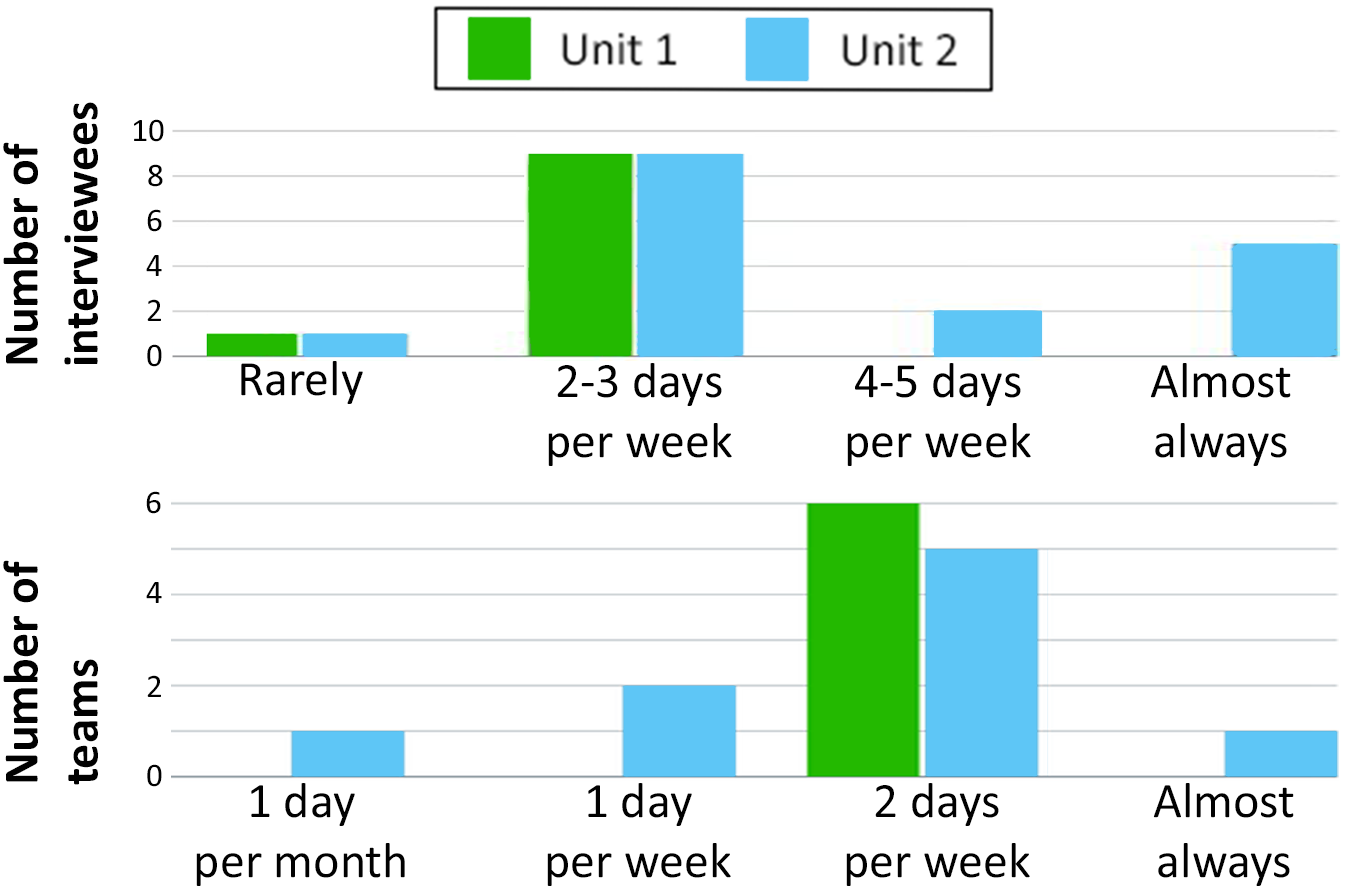}
    \caption{Office presence of the interviewees and teams.}
    \label{fig:presence}
\end{figure}

The office presence of the 27 interviewees and the 15 agile teams is visualised in Figure \ref{fig:presence}. Most interviewees are present at the office minimum two days per week in accordance with the hybrid work policy. Only two interviewees are rarely at the office, due to explicit remote work permissions. All members of the six agile teams in Unit 1 are usually at the office on Tuesdays and Thursdays following the unit implementation of the 2-day office policy. In Unit 2, five agile teams are in the office on the same two days each week. The exact days vary between these teams, as they are agreed on at the team level. One team works together from the office one day per month, and two others one day per week. The ninth team works primarily from the office on all working days.

The majority of the interviewees expressed satisfaction with the two different unit level implementations of the hybrid work policy. Only one interviewee from Unit 2 said that they did not see the point of having two office days, but this was stated in relation to the one-day-per-week team agreement about office co-presence: \qq{I don't really see the point of this second day myself, because if someone or half of the team is working remotely, then we anyway have the meetings via [MS Teams]. [...] If we should be here for two days, then maybe everybody should be here on the same days, but otherwise one day is enough.} --P24 [Team I].

\subsection{Office Workspace and Tools}

The office premises of the company contains meeting rooms of various sizes. The larger rooms are outfitted with conference equipment, i.e., monitors, speakers, and microphones. Some of these rooms are MS Teams equipped \cite{MST}, so they are specifically adapted to meet the needs of hybrid workers. MS Teams is the primary collaboration application for both units. It is used by all the employees for scheduling and hosting meetings. An overview of other tools used to conduct or support meetings is provided in Table \ref{tab:tools}. 

 \begin{table}[b]
\caption{Tools}
\label{tab:tools}
\centering
\begin{tabular}{ p{2.3cm} p{5.7cm} } 
 \toprule
\multicolumn{1}{l}{\textbf{Tool}} &
\multicolumn{1}{l}{\textbf{Meeting Type}} \\
 \midrule
MS Teams &  All \\ 
Jira & Daily, Sprint Planning and Refinement \\ 
Physical whiteboard & Sprint retrospective and Sprint planning \\ 
Digital whiteboard & Sprint retrospective and Sprint planning \\ 
Confluence & Sprint retrospective, Sprint change and CoP's \\ 
Mentimeter poll & Sprint retrospective \\ 
Post-its & Sprint retrospective \\ 
PowerPoint & Operational meetings \\ 
OneNote & PO meetings \\ 
 \bottomrule
\end{tabular}
\end{table}

All of the larger meeting rooms need to be booked in advance, while the smaller meeting rooms can be used if they are available. The smaller meeting rooms can fit 2--4 people, and are primarily only equipped with monitors. Given that the larger meeting rooms need to be booked in advance, the meetings held in these rooms are limited to a fixed time slot: \qq{So [the meeting] usually never goes over the time because the meeting rooms are booked for a specific time. And we have so many teams that every team at some point would need that meeting room, so they kick us out [...] so not all the topics are always covered when we are face-to-face.} --P9 [Team F]. Remote meetings in contrast are not time-constrained in the same manner, which allows for more in-depth discussions when needed. 

Both units have an open space on the company premises where the desks for the employees are spread out. At the time of the data collection, the company used a flexible seating system where employees would book the desks in Flowscape. Despite the free seating, the majority of the teams in both units would almost always sit together at the same desks when they were onsite. Only one team from Unit 1 said that the members were spread out in the open office area when they worked onsite. Two teams have a specific closed-off area to work in, due to security reasons, which is separated from the rest of the teams. In these teams, the members do not need to move to a specific room during meetings, as there are only two teams in this area, and it is large enough that they do not disturb each other. The area they are sitting in is equipped with conference equipment,  making hybrid meetings easy to carry out. The managers from both units also work from a separate closed room, so they carry out most of their meetings in that room when onsite.

\subsection{Unit Meetings}

\begin{figure}[t]
    \centering
    \graphicspath{ {./images/} }   \includegraphics[width=0.42\textwidth]{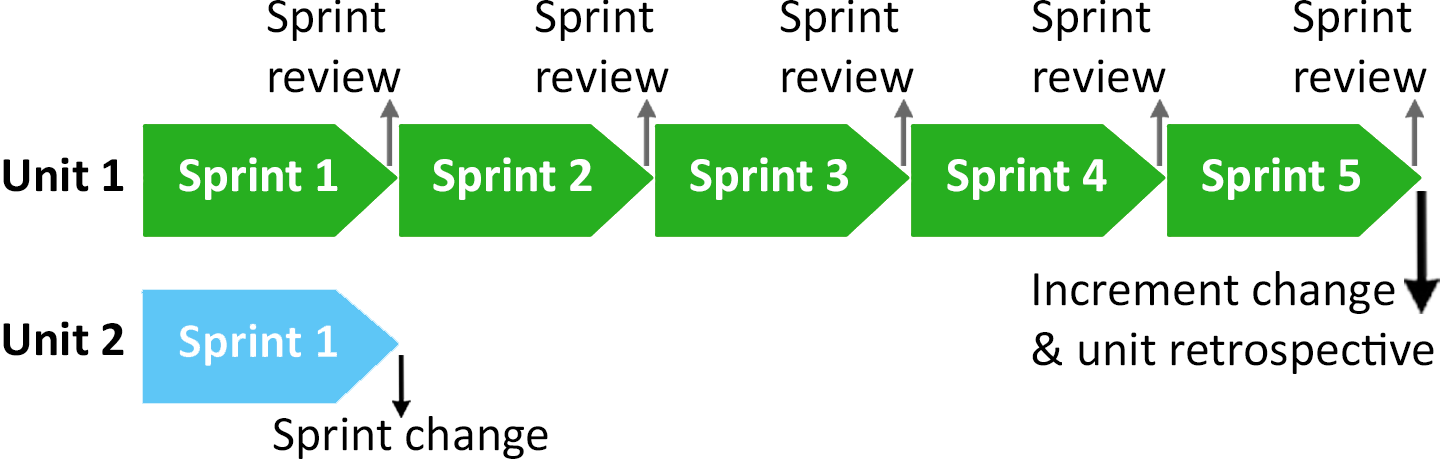}
    \caption{Unit meetings.}
    \label{fig:units}
\end{figure}

The six teams in Unit 1 all follow the same iteration schedule of two-week sprints, with five sprints combined to produce one increment (see Figure \ref{fig:units}). The whole unit meets in a single room at the office every 10 weeks for the \textit{increment change} meeting which lasts 2--3 hours. The work achieved by each team is first presented, usually by the PO, and product management explains what should be achieved in the next increment. Following this, a common \textit{retrospective} is carried out. Here anyone can suggest topics to improve, and each topic is discussed in smaller groups, then presented in plenum. Based on these discussions, action points are made for the whole unit. In addition, remote \textit{sprint review} meetings for all the agile teams in Unit 1 are carried out bi-weekly on Monday evenings. The program manager delivers presentations or demos, and there is the possibility for discussion at the end of the meeting.

All nine agile teams in Unit 2 also follow a mutual iteration schedule with two-week sprints. They have a common remote bi-weekly \textit{sprint change} event for all of the teams on Wednesdays (see Figure \ref{fig:units}). The team coach for each team begins by sharing the work their team completed and key learnings from the previous sprint in plenum. Following this, the participants have smaller discussions on a team level in breakout rooms where they plan and discuss what they will be working on in the coming sprint: \qq{The sprint planning nowadays, the high level one on a project level, the one which is happening on Wednesdays. It's not a planning really. It's more like a celebration [...] it gives a high-level image to everybody of where we are and that would actually explain why there are certain priorities for a certain sprint.} --P13 [Product owner]. Before the meeting, the teams document their key accomplishments, learnings, and work for the next sprint in Confluence, so it is visible to everyone in the unit.

\subsection{Agile Team Meetings}

\begin{figure*}[ht]
    \centering
    \includegraphics[width=\linewidth]{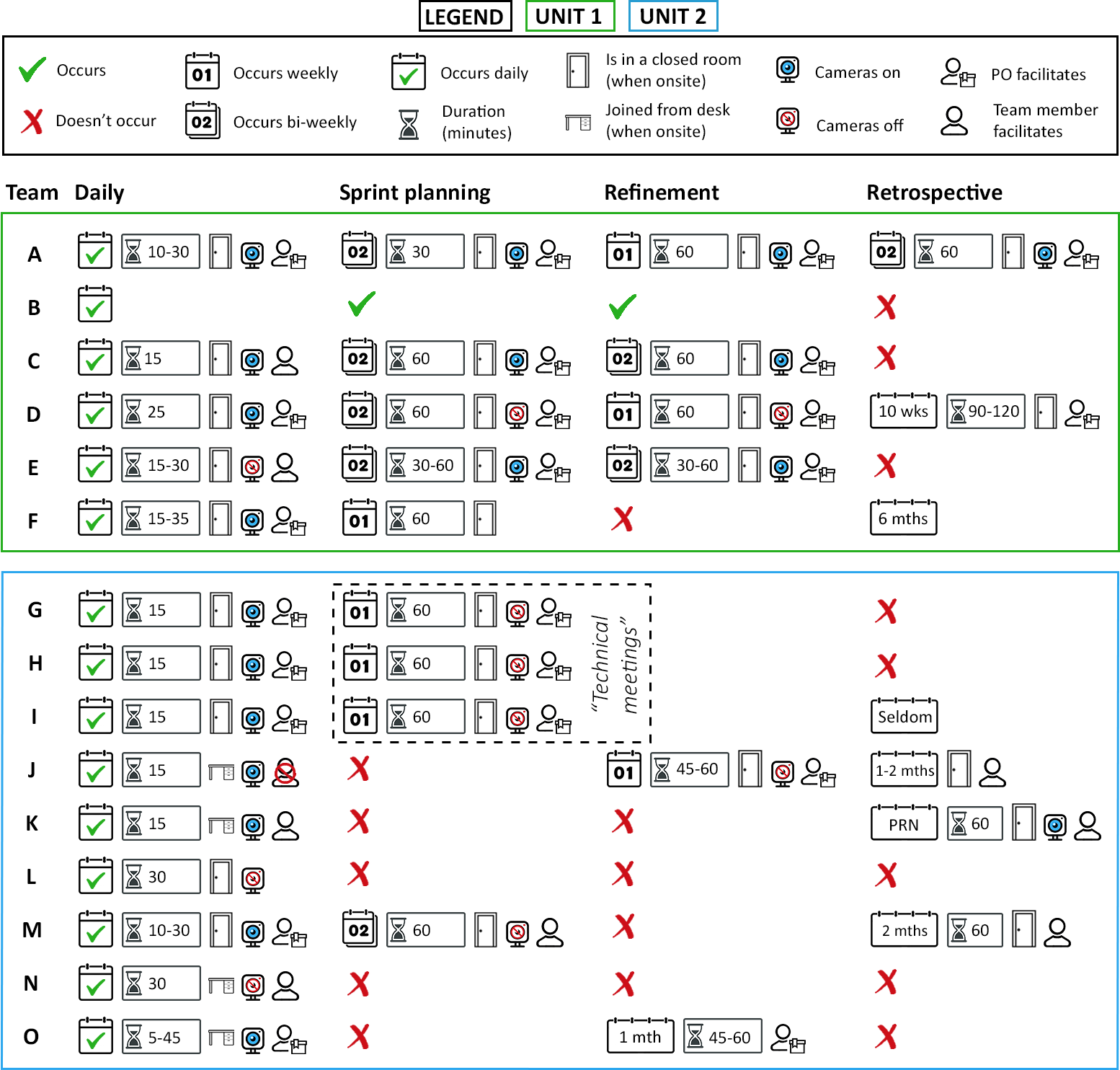}
    \caption{Agile team meetings (PRN = as needed).}
    \label{fig:teammeetings}
\end{figure*}

All the agile teams in the two units held several different recurring meetings for the team members, where the POs often also participated. An overview of the main types of agile team meetings in the two units is provided in Figure \ref{fig:teammeetings}. The occurrence of these meetings varied slightly among teams.  Sprint reviews were not carried out by any of the teams, as these were held during the previously discussed unit meetings.

\subsubsection{Daily meetings} 
\label{sec:daily}
All interviewed agile teams in both units held daily meetings in the morning to discuss their progress and any impediments they encountered. The daily meetings lasted 5--45 minutes, depending on the number of team members and topics to discuss. Because these meetings were carried out daily, the form of the meeting (hybrid, onsite, or remote) always varied. If some team members were onsite, they usually joined the daily from a reserved meeting room. Only four of the teams joined the dailies from their work desks when onsite: \qq{[The team members] are usually joining from their desks. I’m also doing so when I have an office day. Everyone is using headsets to avoid background noise and echo.} --P23 [Team N]. When the daily meeting was fully remote, all the participants joined via MS Teams, and the application was also used during hybrid meetings: \qq{It's [the daily] always hybrid nowadays. Even if we are all in the office, we still go to the same meeting room and we connect to [MS Teams] if needed.} --P18 [Team H]. The dailies were described as being particularly useful meetings by five interviewees. One of these interviewees especially highlighted their importance during days when the teams were not working together at the office: \qq{These couple of days when you are not in the office, you really take the time in the morning to discuss with your team if you have any blockers or if you need any input. That's super important.} --P8 [Team E].

If the daily was remote or hybrid, most of the teams usually had their cameras on during the meeting: \qq{We just noticed that it's much nicer to actually see the person you're talking to [...] there is not a rule from company or anything, it's just in the team. We decided it's better like this.} --P15 [Team G]. Only three of the teams did not use cameras during dailies and the reasons provided by the interviewees for this included for example, poor internet connections for some members, or a lack of need because the whole team was viewing the Jira board via screen-sharing during the meeting: \qq{There's no real benefit [to using cameras]. I'm not staring at the camera anyway [...] I'm looking at my own screen.} --P23 [Team N]. The facilitation of the dailies varied. In seven teams, the PO usually led and facilitated the meeting. Whereas, a team member was responsible for this task in four teams: \qq{the [PO] is not the moderator of the daily anymore, so the team itself took ownership of that meeting} --P4 [Team C]. Only one team did not have anyone in particular who facilitated the dailies.

\subsubsection{Sprint planning and refinement} Four of the agile teams from Unit 1 carried out onsite sprint planning meetings on Tuesdays on a bi-weekly basis, following the unit level sprint review meeting: \qq{When it comes to actually in the team planning the work that we are gonna do, the most important in-person meeting is user story planning, because there the team should sit together and really discuss the feature that they're working on.} --P4 [Team C]. Only one team held sprint planning meetings every week. In this team, the refinement of backlog items was done during the planning meetings, while four of the other teams in the unit held separate weekly or bi-weekly refinement meetings.

In contrast to Unit 1, most of the teams in Unit 2 carried out their planning for the sprints during the previously mentioned sprint change event. Only one team from Unit 2 carried out an additional bi-weekly remote sprint planning meeting specifically for their team: \qq{We sit down as a team and decide and prioritize what should be done and what should we take up next Sprint. And we also get the [PO] involved and as a team then decide the backlog and put the work items for the next sprint in the sprint list [...] it's mostly been on remote days.} --P21 [Team M]. However, three of the teams in Unit 2 had weekly one-hour time slots for \qq{technical meetings}, during which they had discussions around various topics as needed, e.g., planning, features, or customer cases. All the sprint planning meetings lasted 30--60 minutes, and the majority were facilitated by the PO. Only one team had sprint planning meetings that were facilitated by a team member. Jira was the primary tool used during the meetings, but both physical and digital whiteboards were also utilized by the teams.

Similar to Unit 1, one team from Unit 2 carried out weekly backlog refinement meetings, and one other team usually held these meetings monthly. In both units, the refinement meetings lasted 30--60 minutes and the PO always facilitated. When the sprint planning and refinement meetings were hybrid, cameras were used less often than during the dailies, as the teams were focused on the Jira board. In contrast with the dailies, none of the interviewees said that sprint planning or refinement meetings were carried out from their desks when team members were onsite. Only two interviewees did not mention the onsite location of the refinement meetings.

\subsubsection{Retrospectives} Only one team from the two units organized retrospectives specifically for their team on a bi-weekly basis. Whereas, six teams from the two units held retrospectives for their members with varying occurrence, ranging from every 1--6 months, to \qq{very seldom}, or as needed (PRN). The remaining eight teams did not organize their own retrospectives. A reason provided by three of the interviewees for the lack of retrospectives was that the team had been working together for a long time and quickly resolved issues when they arose: \qq{Most of us have been together very long and it's been kind of smooth working in this team. So we don't have that many issues to figure out.} --P15 [Team G]. 

A similar sentiment was expressed by an interviewee from one of the teams that had infrequent retrospectives: \qq{There's a lot of harmony between the developers and the members in the team, because they have the domain knowledge for the past 20 years and they're pretty much in sync about what the objective is, and what the goal is, and hence there is not much argument. That's why within teams there's probably not much need of retrospectives that frequently.} --P9 [Team F]. Only one person from the teams with no retrospectives expressed a potential benefit of these meetings:  \qq{That's something that should be done regularly. If not every sprint, I would say at least every couple of sprints, because you don't really know what to improve on unless you really discuss it.} --P4 [Team C].

All the teams which carried out their own retrospectives did so either onsite from a meeting room, or offsite, and only rarely in hybrid form if, for example, a team member was unable to join in-person. Having the retrospectives as face-to-face meetings was discussed by four of the agile team members as preferable: \qq{We decided to do it on those days when everybody is at the office, or then everybody will come to the office when we book it beforehand.} --P26 [Team J]. The retrospectives lasted 1--2 hours, and if the meeting was hybrid, cameras were always used. Other tools used during the retrospectives included physical and digital whiteboards, post-its, Confluence, and Mentimeter polls.

\subsubsection{Inter-team meetings}
Some of the agile teams which were working on similar features or domains organized joint team meetings to share relevant information among them. For example, teams H and I joined all each other's daily meetings, while teams M and N had a combined daily meeting once per week: \qq{We have a common daily, so both the teams join. So that's on Thursday for both. That's sort of our common information sharing.} --P21 [Team M]. All the participants joined these inter-team dailies via MS Teams because the teams might not always be onsite at the same time. Finally, teams G and K met twice per week to discuss the development features they were working on and how to develop the test environment (not depicted in Figure \ref{fig:teammeetings}).

\subsection{Leadership Meetings}

\subsubsection{High level planning meeting}
A high level planning meeting takes place in Unit 1 in the extended leadership team before the increment change meeting. Here the managers, POs, and specialists define the higher level plans for the coming increment. The meetings are typically carried out in the office because a lot of discussion takes place. 

\subsubsection{Operational meeting}
The extended leadership team from both units have three regular operational meetings per week where ongoing topics are discussed. These meetings are held in the office on Tuesdays and Thursdays and include the management and the POs, as well as some specialists. 

\subsubsection{PO meetings} The POs in Unit 2 have status update reporting meetings every Monday, where they go through all the items they are working on and discuss ideas. They also have a weekly planning meeting on Fridays where they discuss coming features, problems, and delays. Managers from Unit 2 also sometimes join these meetings. All of these meetings are carried out in hybrid form, and most of the POs have cameras on when they are attending remotely. 

\subsubsection{One-on-one meetings}
In both units, 30-minute one-on-one meetings are scheduled once a month between the line managers and the unit employees. The meetings can sometimes happen more often, for example, when new employees join the unit. In Unit 1, these meetings are usually carried out virtually via MS Teams to maintain the privacy of employees and because it is more efficient for the manager. Whereas, in Unit 2 one of the line managers normally organizes the meetings onsite when possible. Cameras are always used by both participants when the one-on-one meetings are virtual.

\subsection{Community Meetings}

In both units, meetings for the various communities of practice (CoP) are held regularly. There are two main types of community meetings: open community meetings where anyone from the unit can suggest a topic and everyone can participate, and closed meetings for role-specific community members (e.g., security leads, architects, and testers). 

\subsubsection{Open CoP meetings} In Unit 1, the open CoP meetings are organized twice per week. A one-hour meeting is held onsite on Thursdays (a unit office day) in a large meeting room, and the other session is a 30-minute remote meeting which is held on Fridays (a unit remote day). These meetings were purposefully placed so attendees can schedule the topic based on the meeting intent. For example, topics that are better for face-to-face, or require discussions, can be scheduled for the office day. In Unit 2, the open CoP meetings are scheduled for specific time slots 2--3 times per week. The meetings are hybrid and most participants join online, but occasionally meeting rooms are booked for them. The meeting invitation shows if a room is booked and contains a link to the Confluence wiki page where employees can write a topic they would like to discuss: \qq{it's up to the team or to the person who can book a CoP session based on the needs and just inform it early enough. And the CoP sessions are in our calendars. [...] And this is something that I really like, to have this weekly based important information flow.} --P8 [Team E]. If no topics are suggested for a given time slot, the meeting does not take place.

There is some overlap between the CoP meetings in Unit 2 and other agile team meetings, in particular the dailies, because the Asian development sites' time zones must be taken into account when scheduling them. Due to this overlap, not all team members can always participate, as discussed by four interviewees: \qq{that's a bit challenging because many teams have their team dailies also at that time [...] people feel that they would need to join the team daily. That's the most important meeting of the team in a day. And then on the other hand, there might be really interesting topics [in the CoP], and that's creating a conflict in a sense that you need to choose [...] but there is not a silver bullet for this.} --P16 [Manager]. To ensure that the information is accessible, the meetings are often recorded so employees can view them later. Some of the agile teams have also decided that the team coach must attend these CoP meetings and share the information with the rest of the team.

\subsubsection{Closed CoP meetings} The closed CoPs in both units are also referred to as ``virtual teams'' and each specific CoP has regular meetings where the community members (e.g., security leads, architects, and testers) have detailed discussions regarding the work in that area: \qq{I like to use the term \textit{virtual team}, because they have their home team, but then they operate in this community where they share experiences. The intent is that each of the community members act as an ambassador towards the community, and then they bring back the best practices to their team.} --P1 [Manager]. 

These meetings are especially important for the specialists who are not connected directly to the agile teams, and to ensure that all of the teams are in sync: \qq{They're big topics, the ones that we are discussing there. And the [CoP] is used as a meeting to align views, because there are quite many teams.} --P10 [Specialist]. The amount and form of closed CoP meetings varies between communities. The architecture community in Unit 1 holds weekly two-hour meetings, alternating between onsite and remote. This alternation addressed a previous challenge voiced by the developers to help ensure that the architects were also available to talk to them when they were onsite: \qq{Other developers that are not part of that meeting were complaining that: Hey, we cannot talk with the [specialists] because they are trapped there in the meeting for two hours. So can you change it a bit? So we have it like this; one week face-to-face, and one week remote.} --P10 [Specialist]. 

The security community in Unit 2 also holds weekly meetings that are typically carried out in hybrid form. A meeting room is often booked at the office but MS Teams is used simultaneously, and community members often join the meeting from their desks. The testing community in Unit 2 had the most meetings, with at least three regular meetings per week. These are also joined by the employees from one of the international sites to ensure that both sites are working in sync. These meetings are carried out in the same form as for the security community and are referred to as ``dailies''.  

Three of the interviewees mentioned in connection to the various closed CoP meetings that cameras are not used as often when people from different sites are joining the meeting, compared to the meetings among members of the Finnish site only. One of the main reasons given for this was poor bandwidth, particularly with the Asian sites. However, the lack of social connection between members may also be a contributing factor: \qq{when we have people who meet less often and are not so much the same community. I noticed many times that people just kind of leave the cameras off, and don't want to share so much of the private area.} --P27 [Manager].

\subsection{Information Sharing Event}
Unit 1 has regular information sharing events. These are carried out onsite as 20-minute informal coffee meetings every Tuesday and Thursday at 14:00. Besides informal information sharing, these events also provide an opportunity to introduce new employees in the unit, and social bonding for all: \qq{whenever we are onsite we have this coffee break. So all the teams go to this common room and have coffee together, chitchat. So it's like a relaxing time of 20 minutes where you can get to know each other more and relax a bit.} --P9 [Team F]. 

\subsection{Meeting Preferences}
In the previous sections, we reported our findings regarding various recurring meetings and how they are organized and carried out. We extend our findings in this section to include some of the preferences voiced by interviewees about meetings in hybrid work in general.

The \textit{intent} of the meeting, i.e., the aim and purpose, affected the choice of how to organize the meeting: six interviewees mentioned that they prefer to have meetings that involve brainstorming onsite, as this allows for more spontaneous discussion and the use of physical whiteboards. Face-to-face meetings were also discussed by six interviewees as useful for other reasons, for example, because it is easier to ask questions: \qq{I think the meetings when we meet face-to-face in the meeting rooms are actually quite efficient, maybe more efficient than the [MS Teams] meetings I would say. It might be easier to raise your questions there than in the [MS Teams] meetings} --P7 [Team D]. 

Two managers and one PO highlighted their preference to schedule virtual meetings involving international attendees on remote working days because the other attendees were also remote. Time zone differences among the sites also contributed to their decision. In addition, attendees can quickly switch between meetings when they are virtual. This can be more efficient when they have either a large number of meetings or back-to-back meetings during the work day, as discussed by two managers and one PO: \qq{I'm just swapping from one [meeting] to the other without any time loss. But if I had to attend to a room, then it takes me five minutes to get from one to the other and then I'm obviously late, so it's much easier to do it like this.} --P13 [Product owner]. 

One interviewee also voiced a preference for attending large information sharing meetings virtually, instead of joining the meeting from a large meeting room at the office, as this provided a better participation experience: \qq{We have this one meeting where there's a lot of people supposed to be there, so it's always cramped. You don't see anything from the back of the meeting room [...]  [When using MS Teams] everyone has their headset on. Everyone has their own screen. Someone is throwing something so you can see it from your own screen, like really close and someone is talking right to your ear. So no information is missed at all. So that is definitely something that's worth mentioning that is good.} --P4 [Team C].

Finally, 18 out of 27 interviewees discussed their preference for using cameras during virtual meetings, describing it as useful or a good practice: \qq{In my teams, everyone is always using the camera [...] that is to me a very good practice, to have these cameras open and show that you are there.} --P11 [Manager]. Only one person said that they did not feel the need to use cameras as is reported in Section \ref{sec:daily} in relation to daily meetings. The remaining eight interviewees did not specifically discuss their preferences around cameras, although three of these interviewees did mention that they used them. One of the interviewees did voice concerns that people possibly keep their cameras off because they are multitasking during meetings. Despite this concern, only three interviewees mentioned that they multi-tasked during virtual meetings.

\section{Discussion}
\label{sec:discussion}

In this section, we discuss our findings, relate them to existing literature, and highlight the key takeaways. We discuss the threats to validity towards the end.

\subsection{Organizing Recurring Meetings} 

A key insight from this study of hybrid work is that different types of recurring meetings in agile software development should be primarily organized onsite or remotely based on the \textit{meeting intent}, i.e., meetings requiring active discussion or brainstorming, such as retrospectives or technical discussions, benefit from onsite attendance, whereas information sharing meetings work well in remote format. For example, a virtual meeting for a larger audience with an agenda of presentations may even provide a better participation experience in comparison to a big onsite meeting room. In this case, a virtual meeting offers good visibility of the content with a clear voice and ensures equal experience for everyone. This insight is similar to the findings in \cite{sporsem2022coordination} regarding large meetings, but differs in regard to retrospectives. While Sporsem and Moe \cite{sporsem2022coordination} found that retrospectives were held online to ensure all team members could participate, the interviewees in our study preferred to hold retrospectives face-to-face.

As most agile meetings are based on active discussion, one might easily think that placing most meetings on office days would be the most efficient. However, our interviewees pointed out problems with that, because people need ``free'' time to meet and converse informally outside the meetings during office days, as discussed also by both Wang et al. \cite{wang2022co} and Sporsem and Moe \cite{sporsem2022coordination}. Especially, specialists and managers, who often have their calendars full of meetings, need to be available also for ad hoc discussions with the agile teams on those days when the team members work at the office.  

We noticed that the layout of the office workspace affected meetings as well, in particular the agile team meetings. Given that employees from both units were expected to work from the office two days per week, the flexible seating system that was in place could be expected to provide benefits, and is discussed by Šmite et al. \cite{smite2022half} as a potential solution for companies that are experiencing only partially used offices, due to hybrid work. However, our findings revealed that the majority of the agile teams preferred fixed seating arrangements and had in practice turned the flexible seating system into fixed seating by always booking the same seats for their team. This allowed them to organize and attend team meetings from their desks, as well as have ad-hoc discussions easily.

\subsubsection*{\textbf{Takeaway}} Organizing numerous back-to-back meetings on office days can hinder informal ad hoc conversations, so we recommend that meetings requiring active discussions or brainstorming are organized onsite, while information sharing meetings can be remote. More research around office workspace design and flexible seating systems is needed, as these can have a direct impact on collaboration and meetings in hybrid work.

\subsection{Community Meetings in Hybrid Work}  

The community of practice (CoP) meetings discussed by the interviewees in this study provide an ideal platform to share knowledge not only inside a team, but also between the teams and across the whole organization. Even though attendance is not always possible for everyone, due to the overlap in meetings, recording the CoP meetings to create a digital footprint and provide non-attendees with access to the information, or having one team member attend the meeting and bring the information back to the rest of their team, ensures that experiences and best practices are continuously shared within the entire community. Thus, these meetings and the strategies used when organizing and carrying them out, provide one solution to the challenge of static and siloed collaboration that was emphasized during remote work in \cite{yang2022effects}. Similarly, meetings like the information sharing coffee events held by Unit 1 provide an opportunity for networking and the strengthening of social ties across the whole unit.  

\subsubsection*{\textbf{Takeaway}} Community meetings and the strategies used when organizing and carrying them out, like video recording, can ensure knowledge is shared within organizations, help strengthen social ties, and prevent siloed collaboration in hybrid work.

\subsection{Using Cameras in Meetings}  
 
The use of cameras during remote and hybrid meetings is a topic that has been discussed and studied especially during the Covid-19 pandemic \cite{rodeghero2021please, shockley2021fatiguing}. The majority of our interviewees found that having cameras on during hybrid and remote meetings was useful and a good practice. At the team level, the use of cameras was discussed as particularly beneficial for meetings like the dailies and retrospectives, where participants discuss with one another. In contrast, during more technical meetings, like backlog refinement meetings, everyone looks at the shared screen (e.g., Jira board), therefore having cameras on was not felt that useful. Additionally, in cross-site meetings that involve people who meet less often, they tend to more easily leave their cameras off. More research in this area is needed, as using cameras may be beneficial to help strengthen networks between people working in different sites. Finally, despite the concern voiced by one interviewee regarding possible multitasking when cameras were kept off, this behavior was not prevalent in our results, which stands in contrast to findings from previous research \cite{cao2021large}.  

\subsubsection*{\textbf{Takeaway}} Cameras can be beneficial in remote and hybrid meetings that require active discussion between participants, but may not be as useful during more technical meetings. The use of cameras therefore requires further attention to learn their advantages or disadvantages.

\subsection{Threats to Validity}
We report threats to validity per the guidelines by Runeson and Höst \cite{runeson2009guidelines}. Our case study is exploratory and does not concern causal relationships, therefore, \textit{internal validity} aspects do not apply \cite{yin2018case}. 

\subsubsection{Construct Validity}
Our study is not prone to the threats of construct validity because our interview guide contained a substantial amount of questions related to collaboration in hybrid work. Additionally, certain questions were explicit in querying about the meetings in hybrid work. All the interviewers and the study participants are knowledgeable about agile development process models and practices, and the different types of meetings in agile, which alleviates the possibility of a different understanding of the construct between the participants and the interviewers. The feedback sessions also strengthened the construct validity, although some limitations do exist, given that we could not confirm all interviewees were present at the sessions. A draft of this paper was also shared with the company managers for validation, and they confirmed that the results were accurate.

\subsubsection{External Validity}
The external validity of case studies is limited to the studied context \cite{runeson2009guidelines}. Our findings related to meetings are yet generalizable to similar contexts, i.e., companies that apply hybrid work policies and models in large-scale agile software development, as well as companies that seek to transform their workspaces to facilitate a collaborative environment for hybrid work. Moreover, our sample was not limited to the members of agile teams, but included multiple roles, with varied years of experience in agile software development. 

\subsubsection{Reliability}
We implemented multiple measures to enhance the reliability of our study. Our integrated coding approach promotes the possibility of developing a similar codebook by other researchers. The iterative coding approach and joint meetings among all the authors to discuss the codebook further improved the reliability. We publicly share the interview guide and codebook with descriptions of all levels of coding to further contribute to the reliability aspects. Thus, decreasing the subjectivity in data analysis.

\section{Conclusion and Future Directions}
This study offers qualitative insights into how recurring meetings are organized and carried out in hybrid work, in a large-scale agile environment. We performed a single case study in the Finnish R\&D site of Ericsson with a ``2 days per week at the office'' policy. 27 participants including agile team members from 15 different teams, managers, product owners and specialists participated in the semi-structured interview data collection. We observed that sprint planning and retrospectives are preferably organized and carried out onsite by the agile teams because they require active discussions, while daily meetings are typically hybrid. In addition, organizing a hybrid participation possibility for all meetings is often necessary to include everyone. 

\textit{For practitioners:} It is better to arrange meetings requiring active discussions or brainstorming onsite, whereas, information sharing meetings can be remote. Office days should not be booked with numerous back-to-back meetings because this can limit informal, yet important, ad hoc discussions. Community meetings can contribute to knowledge sharing within organizations, and help strengthen social ties and prevent siloed collaboration in hybrid work. Additionally, the use of cameras is recommended for small discussion-oriented remote and hybrid meetings.

\textit{For future research:} Studying the dynamics of ad hoc meetings in hybrid work, and understudied meetings like technical meetings, can offer more useful insights. In companies with varying hybrid work models, future research could investigate how having both predefined and flexible office days impacts inter-team collaboration in meetings. The use of cameras in remote and hybrid meetings also requires further attention to learn their advantages or disadvantages.

\section*{Acknowledgment}
We would like to thank Ericsson for their engagement in our research, and the Finnish Work Environment Fund for funding
the research.

\bibliographystyle{IEEEtran}
\bibliography{bibliography}

\begin{thebibliography}{10}
\providecommand{\url}[1]{#1}
\csname url@samestyle\endcsname
\providecommand{\newblock}{\relax}
\providecommand{\bibinfo}[2]{#2}
\providecommand{\BIBentrySTDinterwordspacing}{\spaceskip=0pt\relax}
\providecommand{\BIBentryALTinterwordstretchfactor}{4}
\providecommand{\BIBentryALTinterwordspacing}{\spaceskip=\fontdimen2\font plus
\BIBentryALTinterwordstretchfactor\fontdimen3\font minus \fontdimen4\font\relax}
\providecommand{\BIBforeignlanguage}[2]{{%
\expandafter\ifx\csname l@#1\endcsname\relax
\typeout{** WARNING: IEEEtran.bst: No hyphenation pattern has been}%
\typeout{** loaded for the language `#1'. Using the pattern for}%
\typeout{** the default language instead.}%
\else
\language=\csname l@#1\endcsname
\fi
#2}}
\providecommand{\BIBdecl}{\relax}
\BIBdecl

\bibitem{smite2022half}
D.~Šmite, N.~B. Moe, A.~Tkalich, G.~K. Hanssen, K.~Nydal, J.~N. Sandb{\ae}k, H.~W. Aamo, A.~O. Hagaseth, S.~A. Bekke, and M.~Holte, ``Half-empty offices in flexible work arrangements: Why are employees not returning?'' in \emph{International Conference on Product-Focused Software Process Improvement}.\hskip 1em plus 0.5em minus 0.4em\relax Springer, 2022, pp. 252--261.

\bibitem{conboy2023future}
K.~Conboy, N.~B. Moe, V.~Stray, and J.~H. Gundelsby, ``The future of hybrid software development: Challenging current assumptions,'' \emph{IEEE Software}, vol.~40, no.~02, pp. 26--33, 2023.

\bibitem{yang2022effects}
L.~Yang, D.~Holtz, S.~Jaffe, S.~Suri, S.~Sinha, J.~Weston, C.~Joyce, N.~Shah, K.~Sherman, B.~Hecht \emph{et~al.}, ``The effects of remote work on collaboration among information workers,'' \emph{Nature human behaviour}, vol.~6, no.~1, pp. 43--54, 2022.

\bibitem{miller2021your}
C.~Miller, P.~Rodeghero, M.-A. Storey, D.~Ford, and T.~Zimmermann, ``How was your weekend? software development teams working from home during covid-19,'' in \emph{2021 IEEE/ACM 43rd International Conference on Software Engineering (ICSE)}.\hskip 1em plus 0.5em minus 0.4em\relax IEEE, 2021, pp. 624--636.

\bibitem{muller2023challenges}
K.~M{\"u}ller, C.~Koch, D.~Riehle, M.~Stops, and N.~Harutyunyan, ``Challenges of working from home in software development during covid-19 lockdowns,'' \emph{ACM Transactions on Software Engineering and Methodology}, vol.~32, no.~5, pp. 1--41, 2023.

\bibitem{jaspan2023developer}
C.~Jaspan and C.~Green, ``Developer productivity for humans, part 2: Hybrid productivity,'' \emph{IEEE Software}, vol.~40, no.~2, pp. 13--18, 2023.

\bibitem{tkalich2023pair}
A.~Tkalich, N.~B. Moe, N.~H. Andersen, V.~Stray, and A.~M. Barbala, ``Pair programming practiced in hybrid work,'' in \emph{2023 ACM/IEEE International Symposium on Empirical Software Engineering and Measurement (ESEM)}.\hskip 1em plus 0.5em minus 0.4em\relax IEEE, 2023, pp. 1--7.

\bibitem{de2023post}
R.~de~Souza~Santos, G.~Adisaputri, and P.~Ralph, ``Post-pandemic resilience of hybrid software teams,'' in \emph{2023 IEEE/ACM 16th International Conference on Cooperative and Human Aspects of Software Engineering (CHASE)}.\hskip 1em plus 0.5em minus 0.4em\relax IEEE, 2023, pp. 1--12.

\bibitem{khanna2024hybrid}
D.~Khanna, E.~L. Christensen, S.~Gosu, X.~Wang, and M.~Paasivaara, ``Hybrid work meets agile software development: A systematic mapping study,'' in \emph{Proceedings of the 2024 IEEE/ACM 17th International Conference on Cooperative and Human Aspects of Software Engineering}, 2024, pp. 57--67.

\bibitem{agilereport}
\BIBentryALTinterwordspacing
Digital.ai. (2023) 17th state of agile report. [Online]. Available: \url{https://digital.ai/resource-center/analyst-reports/state-of-agile-report/}
\BIBentrySTDinterwordspacing

\bibitem{beck2001agile}
\BIBentryALTinterwordspacing
K.~Beck, M.~Beedle, A.~Van~Bennekum, A.~Cockburn, W.~Cunningham, M.~Fowler, J.~Grenning, J.~Highsmith, A.~Hunt, R.~Jeffries \emph{et~al.} (2001) The agile manifesto. [Online]. Available: \url{https://agilemanifesto.org/principles.html}
\BIBentrySTDinterwordspacing

\bibitem{smite2022future}
D.~Šmite, E.~L. Christensen, P.~Tell, and D.~Russo, ``The future workplace: Characterizing the spectrum of hybrid work arrangements for software teams,'' \emph{IEEE software}, vol.~40, no.~2, pp. 34--41, 2022.

\bibitem{stray2016daily}
V.~Stray, D.~I. Sj{\o}berg, and T.~Dyb{\aa}, ``The daily stand-up meeting: A grounded theory study,'' \emph{Journal of Systems and Software}, vol. 114, pp. 101--124, 2016.

\bibitem{stray2018daily}
V.~Stray, N.~B. Moe, and D.~I. Sjoberg, ``Daily stand-up meetings: start breaking the rules,'' \emph{IEEE software}, vol.~37, no.~3, pp. 70--77, 2018.

\bibitem{rodeghero2021please}
P.~Rodeghero, T.~Zimmermann, B.~Houck, and D.~Ford, ``Please turn your cameras on: Remote onboarding of software developers during a pandemic,'' in \emph{2021 IEEE/ACM 43rd International Conference on Software Engineering: Software Engineering in Practice (ICSE-SEIP)}.\hskip 1em plus 0.5em minus 0.4em\relax IEEE, 2021, pp. 41--50.

\bibitem{cao2021large}
H.~Cao, C.-J. Lee, S.~Iqbal, M.~Czerwinski, P.~N. Wong, S.~Rintel, B.~Hecht, J.~Teevan, and L.~Yang, ``Large scale analysis of multitasking behavior during remote meetings,'' in \emph{Proceedings of the 2021 CHI Conference on Human Factors in Computing Systems}, 2021, pp. 1--13.

\bibitem{buyukguzel2023progressivity}
S.~Buyukguzel and R.~Mitchell, ``Progressivity in hybrid meetings: Daily scrum as an enabling constraint for a multi-locational software development team,'' \emph{Computer Supported Cooperative Work (CSCW)}, pp. 1--34, 2023.

\bibitem{buyukguzel2023spatial}
S.~B{\"u}y{\"u}kg{\"u}zel and U.~Balaman, ``The spatial organization of hybrid scrum meetings: A multimodal conversation analysis study,'' \emph{Discourse \& Communication}, vol.~17, no.~3, pp. 253--277, 2023.

\bibitem{sporsem2022unscheduled}
T.~Sporsem, A.~F. Strand, and G.~K. Hanssen, ``Unscheduled meetings in hybrid work,'' \emph{IEEE Software}, vol.~40, no.~2, pp. 42--49, 2022.

\bibitem{sporsem2022coordination}
T.~Sporsem and N.~B. Moe, ``Coordination strategies when working from anywhere: a case study of two agile teams,'' in \emph{International Conference on Agile Software Development}.\hskip 1em plus 0.5em minus 0.4em\relax Springer, 2022, pp. 52--61.

\bibitem{wang2022co}
Z.~Wang, Y.-H. Chou, K.~Fathi, T.~Schimmer, P.~Colligan, D.~Redmiles, and R.~Prikladnicki, ``Co-designing for a hybrid workplace experience in software development,'' \emph{IEEE software}, vol.~40, no.~2, pp. 50--59, 2022.

\bibitem{yin2018case}
R.~K. Yin, \emph{Case Study Research and Applications: Design and Methods}.\hskip 1em plus 0.5em minus 0.4em\relax Sage Thousand Oaks, CA, 2018, vol.~6.

\bibitem{runeson2009guidelines}
P.~Runeson and M.~H{\"o}st, ``Guidelines for conducting and reporting case study research in software engineering,'' \emph{Empirical software engineering}, vol.~14, pp. 131--164, 2009.

\bibitem{brereton2008using}
P.~Brereton, B.~Kitchenham, D.~Budgen, and Z.~Li, ``Using a protocol template for case study planning,'' in \emph{12th International Conference on Evaluation and Assessment in Software Engineering (EASE) 12}, 2008, pp. 1--8.

\bibitem{ericsson}
\BIBentryALTinterwordspacing
Ericsson. (©1994--2024) Ericsson in finland. [Online]. Available: \url{https://www.ericsson.com/en/about-us/company-facts/ericsson-worldwide/finland}
\BIBentrySTDinterwordspacing

\bibitem{paasivaara2018large}
M.~Paasivaara, B.~Behm, C.~Lassenius, and M.~Hallikainen, ``Large-scale agile transformation at ericsson: a case study,'' \emph{Empirical Software Engineering}, vol.~23, pp. 2550--2596, 2018.

\bibitem{hallikainen2011experiences}
M.~Hallikainen, ``Experiences on agile seating, facilities and solutions: multisite environment,'' in \emph{2011 IEEE Sixth International Conference on Global Software Engineering}.\hskip 1em plus 0.5em minus 0.4em\relax IEEE, 2011, pp. 119--123.

\bibitem{braun2021one}
V.~Braun and V.~Clarke, ``One size fits all? what counts as quality practice in (reflexive) thematic analysis?'' \emph{Qualitative research in psychology}, vol.~18, no.~3, pp. 328--352, 2021.

\bibitem{MST}
\BIBentryALTinterwordspacing
Microsoft. (©2024) Microsoft teams rooms. [Online]. Available: \url{https://www.microsoft.com/en-us/microsoft-teams/microsoft-teams-rooms}
\BIBentrySTDinterwordspacing

\bibitem{shockley2021fatiguing}
K.~M. Shockley, A.~S. Gabriel, D.~Robertson, C.~C. Rosen, N.~Chawla, M.~L. Ganster, and M.~E. Ezerins, ``The fatiguing effects of camera use in virtual meetings: A within-person field experiment.'' \emph{Journal of Applied Psychology}, vol. 106, no.~8, p. 1137, 2021.

\end{thebibliography}

\end{document}